# Nonplanar Chiral Metamaterials with Negative Index

Bingnan Wang,[1] Jiangfeng Zhou,[1] Thomas Koschny,[1,2] and Costas M. Soukoulis[1,2,*]

[1]*Ames Laboratory and Department of Physics and Astronomy, Iowa State University, Ames, Iowa 50011, USA*
[2]*Institute of Electronic Structure and Laser, FORTH, and Department of
Materials Science and Technology, University of Crete, Heraklion, Crete, Greece*

We demonstrate experimentally and numerically that non-planar chiral metamaterials give giant optical activity, circular dichroism and negative refractive index. The transmission, reflection and the retrieval results of the experiments agree pretty well with the simulations. This is an important step toward the design and fabrication of three-dimensional isotropic chiral metamaterials.

Recently, a lot of experimental work of chiral metamaterials (CMs) fabricated by planar technologies, has been published[1–7]. The interest is due to the strong optical activity, circular dichroism and to the predictions[8–10] that CMs can be used to achieve negative refraction. Theoretical studies also show properties such as focusing of circularly polarized waves with a chiral medium slab[11,12], and negative reflection in a strong chiral medium[13]. The strong optical activity and circular dichroism in planar chiral metamaterials have been studied experimentally by several groups since 2003[1–4]. Fabricated planar CMs[5] that give negative refractive index was published in 2009.

Similar to metamaterials designed for linear polarized waves, CMs are also periodic arrangements of artificial structures. These artificial structures in CMs are chiral so that cross-coupling between the magnetic and electric fields happens at the resonance. The cross-coupling is described by the chirality parameter $\kappa$ so that the constitutive relations of a chiral medium is given by

$$\mathbf{D} = \epsilon_0 \epsilon_r \mathbf{E} + i\kappa \sqrt{\mu_0 \epsilon_0} \mathbf{H} \quad (1)$$
$$\mathbf{B} = \mu_0 \mu_r \mathbf{H} - i\kappa \sqrt{\mu_0 \epsilon_0} \mathbf{E} \quad (2)$$

where $\epsilon_r$ and $\mu_r$ are the relative permittivity and permeability of the chiral medium, $\epsilon_0$ and $\mu_0$ are the permittivity and permeability of vacuum, respectively.

The eigen-solutions of the electromagnetic (EM) waves in chiral media are the right-handed circularly polarized (RCP, +) wave and the left-handed circularly polarized (LCP, -) wave[14]. Due to the rotational asymmetry, the polarization plane of a linearly polarized wave will rotate when it passed through a chiral medium, introducing optical activity. Also, the RCP and LCP waves interact with the chiral particles differently and are absorbed to a different extent, causing circular dichroism. Moreover, the presence of $\kappa$ causes the difference of refractive index of RCP ($n_+$) and LCP ($n_-$) waves. $n_\pm = n \pm \kappa$, where $n = \sqrt{\epsilon_r \mu_r}$. When $|\kappa|$ is large enough, either $n_+$ or $n_-$ becomes negative. While in conventional metamaterials [15], both negative $\epsilon_r$ and $\mu_r$ are required to get negative $n$, neither $\epsilon$ nor $\mu$ needs to be negative in CMs. This makes negative refraction easier to achieve and offers simpler designs of metamaterials.

Although many proposed models for CMs are three-dimensional (3D), most of the experimental results reported so far are planar structures. Planar structures are easier to fabricate and they have strong optical activity and circular dichroism. Bi-layer rosette-shaped[2,5] and cross wires[7] structures were recently studied in the microwave regime. The patterns are physically separated on each layer and the patterns on the second layer are rotated by an angle. This gives an extremely strong rotation, which is, in terms of rotary power per wavelength, five orders of magnitude longer than a gyrotropic quartz crystal. Recently, negative effective refractive index was demonstrated by parameter retrieval from transmission and reflection measurements[5–7].

While the thicknesses of the planar CMs are much smaller than the working wavelengths, the size of the patterns in plane is usually comparable with half-wavelength [5,7]. Moreover, in effective medium theory, homogeneity of a metamaterial is assumed to calculate the macroscopic parameters. However, some of the proposed planar CMs are not symmetric in the propagation direction[6]. This causes differences in transmission and reflection from opposite directions[16] and ambiguity in the calculation of effective parameters. So these planar CMs are not suitable to build 3D bulk CMs.

In this letter we demonstrate experimentally and numerically the negative refraction in a non-planar CM. We observe strong cross coupling between the magnetic and electric fields by the two coupled split-ring resonators (SRRs) on two sides of the substrates at the chiral resonance frequency. The negative refractive index of CMs arise from this strong chirality. A retrieval procedure[5] adopting uniaxial bi-anisotropic model have been developed to calculate the effective parameters of our CM. Compared with other planar chiral structures, our non-planar CM is symmetric along the propagation direction and much smaller than the working wavelength in all dimensions. So it is a good candidate for isotropic 3D CMs.

The chiral structure of our studies is formed by two identical SRRs separated by a dielectric substrate and interconnected by vias, as shown in Fig. 1(a). The chiral SRRs are arranged in square lattices to form a 2D slab. Fig. 1(b) shows a unit cell of such a 2D array. In this arrangement, each of the chiral SRRs is shared by the two neighboring cells (see Fig. 1(c) for a photo of the fabricated CM). When the structure is excited by external field, both magnetic dipole and electric dipole exist and they are both parallel to the axis of the SRRs. While the induced current around in the SRRs gives the magnetic dipole, electric charges accumulate on the two SRRs with opposite signs, which introduce a strong electric field between the top and bottom SRRs and give the electric dipole in the same direction as the magnetic dipole(see Fig. 2). A electric dipole can excite a magnetic dipole in a similar way.

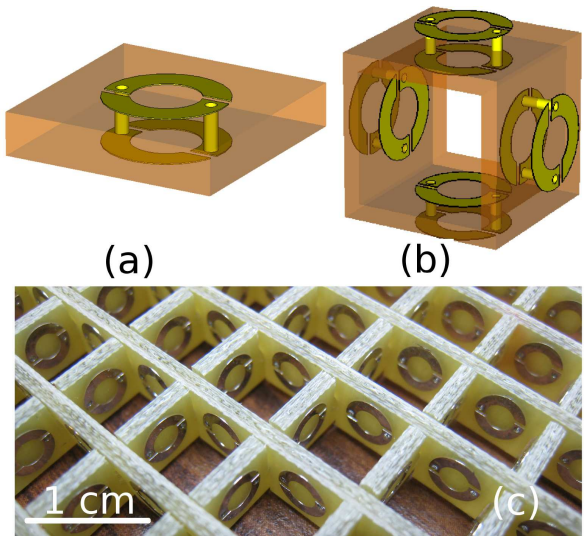

FIG. 1: (color online) (a) The structure of the chiral SRR. Please see text for dimensions. (b) A unit cell of the 2D CM, formed by four interlocked chiral-SRRs. (c) A photo of the fabricated CM.

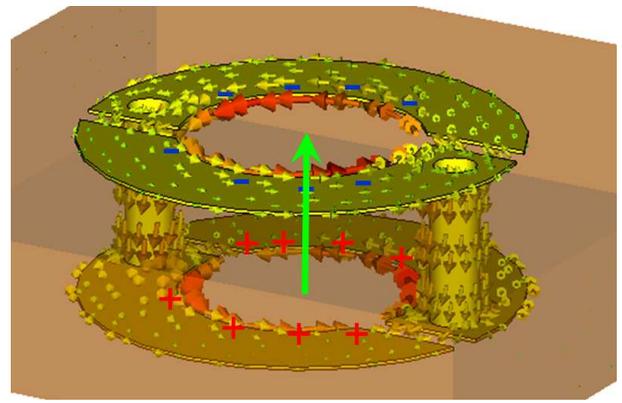

FIG. 2: (color online) The surface current distribution of the SRR near resonance. The circulating current forms a magnetic dipole, causes the opposite charge accumulation on the bottom SRR (+) and the top SRR (-) and forms an extra electric dipole (see green straight arrow).

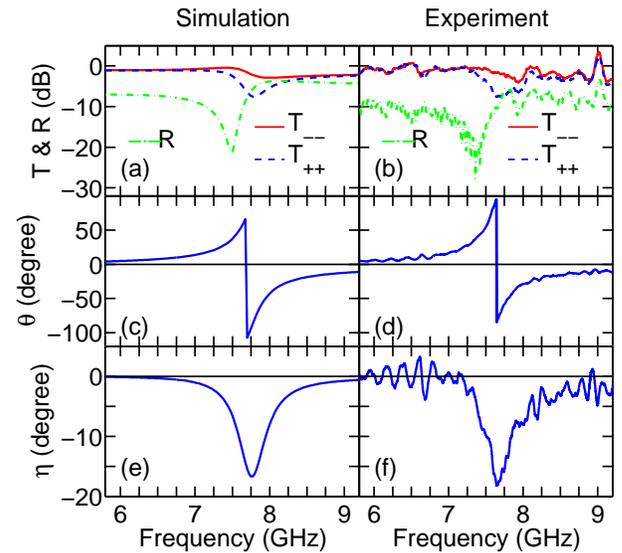

FIG. 3: (color online) The transmission and reflection amplitudes for LCP and RCP (top panel), the azimuth rotation angle for linearly polarized waves (middle panel) and the ellipticity (bottom panel). Simulation results are shown on the left column and experiment results are shown on right column.

This chiral design was proposed [17–19] to develop 3D isotropic CMs [19]. By arranging the chiral SRRs in cubic (fcc) Bravais lattices, a bi-isotropic medium can be obtained. By calculating the susceptibility parameters, including the chiral parameter $\kappa$, by Lorentz local field theory, the medium is shown to provide negative refraction over a frequency band. In this letter, we study a single layer of the CM made of chiral SRRs and demonstrate both numerically and experimentally the strong optical activity and circular dichroism, and negative refraction. This is an important step toward the design and characterizations of 3D isotropic CMs.

The chiral structures are fabricated on the two sides of FR-4 printed circuit boards (PCBs), with a relative dielectric constant of $\epsilon_r = 3.76$ with loss tangent 0.0186 and thickness of 1.6 mm. The metal structures are built on copper with a thickness of 36 $\mu$m. The SRRs are 2-gap split rings with an identical gap width of 0.3 mm. The inner radius of the rings is 1.25 mm and the outer radius is 2.25 mm. The SRRs on opposite sides of the board are connected by vias with a diameter of 0.5 mm. The distance between adjacent rings is 8 mm. The sample is fabricated by 50 strips with 26 cells each, which are then interlocked to form a slab of size 208 mm × 208 mm of the chiral metamaterial. The picture of the sample is shown in Fig. 1(c).

A vector network analyzer (Agilent E8364B) and a pair of standard gain horn antennas are used to perform the transmission and reflection measurements. Since the signals from the horn antennas are linearly polarized, four transmission components $T_{xx}$, $T_{xy}$, $T_{yx}$ and $T_{yy}$ and four reflection components $R_{xx}$, $R_{xy}$, $R_{yx}$ and $R_{yy}$ are measured[20] in our experiments. Then, the circular transmission coefficients[21], $T_{++}$, $T_{-+}$, $T_{+-}$ and $T_{--}$, and reflection coefficients $R_{++}$, $R_{-+}$, $R_{+-}$ and $R_{--}$ are converted from the linear coefficients[7].

Fig. 3(a) and (b) show the simulated and measured transmission coefficients as a function of frequency. There is obvious difference (about 5 dB) in $T_{++}$ and $T_{--}$ around the resonance. The reflection coefficients $R_{+-} = R_{-+} = R$, since the impedances for RCP and LCP waves are identical. The cross coupling transmission $T_{-+}$ and $T_{+-}$ are negligible and they are not shown. While the simulation curves are smooth off the resonance, there are a lot of noise in the experimental curves. This is mainly due to the scattering from the surroundings and multiple reflections between the horn antennas and the sample. Also, the resonance frequency in experiment is about 0.1 GHz lower than the resonance frequency in simulation. This is not surprising since the parameters, particularly the dielectric constant of the PCB, may not be accurate in our frequency

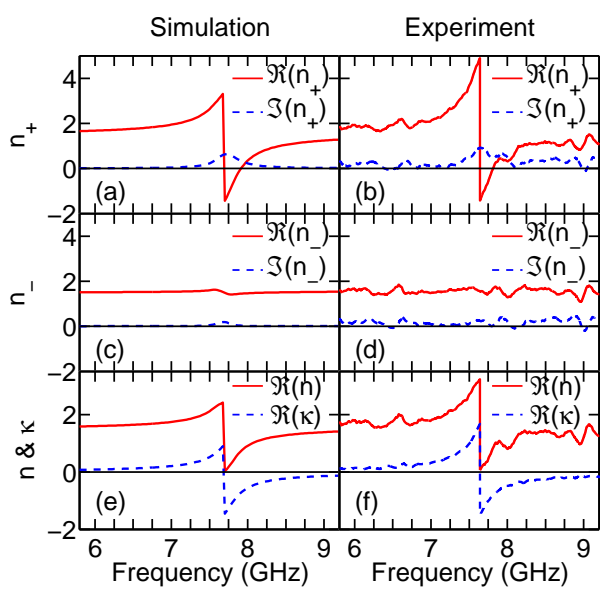

FIG. 4: (color online) The retrieved effective parameters of the chiral metamaterial. The effective index of refraction $n$ together with the chiral parameter $\kappa$ (top panel), the index of refraction for the two circularly polarized waves $n_+$ (middle panel) and $n_-$ (bottom panel). Simulation results are shown on the left column and experiment results are shown on the right column.

range.

Use the standard definition for the polarization azimuth rotation, $\theta = \frac{1}{2}[\arg(T_{++}) - \arg(T_{--})]$, we calculated the polarization changes of a linearly polarized waves incident on our chiral structure. The ellipticity, $\eta = \frac{1}{2}\sin^{-1}\left(\frac{|T_{++}|^2 - |T_{--}|^2}{|T_{++}|^2 + |T_{--}|^2}\right)$, which is a measure of circular dichroism, is also calculated. The simulated and measured azimuth rotation $\theta$, and the ellipticity $\eta$, are presented in Fig. 3 (c)-(f), respectively. At the resonance, the azimuth rotation reaches a maximum of around $-100°$. Meanwhile, the ellipticity is also the largest ($-17°$), meaning a linearly polarized wave is strongly distorted and becomes elliptical.

Fig. 4 shows the retrieved effective parameters of the CM. We see that the effective refraction index for RCP wave $n_+$ shows strong response at the resonance (~7.7 GHz) and goes negative above the resonance (Fig. 4(a) and (b)) while $n_-$ changes only slightly and is positive in all the frequency range (Fig. 4(c) and (d)). The $\epsilon_r$ and $\mu_r$ are not both negative in the same frequency (not shown), which is the signature of conventional negative index metamaterials. $\epsilon_r$ is positive and $\mu_r$ is negative at the resonance frequency 7.7 GHz. The negative index here is due to a relatively large chiral parameter $\kappa$, and a small value of $n$ at the resonance (see Fig. 4(e) and (f)). The negative index has a minimum of around 1.7 with a figure of merit about 2. The peak of the imaginary part of the index of refraction comes from the strong absorption in the PCB substrate at the resonance. The loss can be reduced significantly when a low-loss substrate can be used.

In summary, the electromagnetic properties of the non-planar CM composed of chiral SRRs have been studied. Very strong optical activity, as well as circular dichroism can be achieved with the metamaterial. Moreover, negative index of refraction is obtained by both numerical simulations and experimental measurement. A 3D bulk medium can be fabricated by stacking layers of this CM together. Eventually 3D isotropic CMs, with negative refraction due to chirality, can be obtained.

Work at Ames Laboratory was supported by the Department of Energy (Basic Energy Sciences) under contract No. DE-AC02-07CH11358. This work was partially supported by the Department of Navy, office of the Naval Research (Grant No. N00014-07-1-0359), European Community FET project PHOME (Contract No. 213390) and AFOSR under MURI grant (FA 9550-06-1-0337).

* Electronic address: soukoulis@ameslab.gov
[1] A. Papakostas, A. Potts, D. M. Bagnall, S. L. Prosvirnin, H. J. Coles, and N. I. Zheludev, *Phys. Rev. Lett.* **90**, 107404 (2003).
[2] A. V. Rogacheva, V. A. Fedotov, A. S. Schwanecke, and N. I. Zheludev, *Phys. Rev. Lett.* **97**, 17740 (2006).
[3] M. Kuwata-Gonokami, N. Saito, Y. Ino, M. Kauranen, K. Jefimovs, T. Vallius, J. Turunen, and Y. Svirko, *Phys. Rev. Lett.* **95**, 227401 (2005).
[4] M. Decker, M. W. Klein, M. Wegener, and S. Linden, *Opt. Lett.* **32**, 856 (2007).
[5] E. Plum, J. Zhou, J. Dong, V. A. Fedotov, T. Koschny, C. M. Soukoulis, and N. I. Zheludev, *Phys. Rev. B* **79**, 035407 (2009).
[6] S. Zhang, Y. Park, J. Li, X. Lu, W. Zhang, and X. Zhang, *Phys. Rev. Lett.* **102**, 023901 (2009).
[7] J. Zhou, J. Dong, B. Wang, Th. Koschny, M. Kafesaki, and C. M. Soukoulis, *Phys. Rev. B* (to be published).
[8] S. Tretyakov, I. Nefedov, A. Sihvola, S. Maslovski, and C. Simovski, *J. Electromagnetic Waves Applications* **17**, 695 (2003).
[9] J. B. Pendry, *Science* **306**, 1353, 2004.
[10] S. Tretyakov, A. Sihvola, and L. Jylh, *Photonics and Nanostructures - Fundamentals and Applications* **3**, 107 (2005).
[11] C. Monzon and D. W. Forester, *Phys. Rev. Lett.* **95**, 123904 (2005).
[12] Y. Jin and S. He, *Opt. Express* **13**, 4974 (2005).
[13] C. Zhang and T. J. Cui, *App. Phys. Lett.* **91**, 194101 (2007)
[14] J. A. Kong, *Electromagnetic wave theory* (Cambridge, 2000)
[15] C. M. Soukoulis, S. Linden, and M. Wegener, *Science* **315**, 47 (2007).
[16] V. A. Fedotov, P. L. Mladyonov, S. L. Prosvirnin, A. V. Rogacheva, Y. Chen and N. I. Zheludev, *Phys. Rev. Lett.* **97**, 167401 (2006)
[17] J. D. Baena, L. Jelinek, and R. Marqués, *Phys. Rev B* **76**, 245115 (2007).
[18] R. Marqués, L. Jelinek, and F. Mesa, *Microwave Opt. Technology Lett.* **49**, 2606 (2002).
[19] L. Jelinek, R. Marqués, F. Mesa, and J. D. Baena, *Phys. Rev. B* **77**, 205110 (2008).

[20] In these notations, the first and second subscript indicates the output and input signal polarization, respectively. For example, $T_{xy} = E_x^t/E_y^i$, where $E_y^i$ is the input $y$–polarized electric field and $E_x^t$ is the transmitted $x$–polarized electric field.

[21] Similarly, $T_{++} = E_+^t/E_+^i$, where $E_+^i$ is the input RCP electric field and $E_+^t$ is the transmitted RCP electric field. $R_{-+} = E_-^r/E_+^i$, where $E_-^r$ is the reflected LCP electric field. Note that the RCP/LCP wave becomes LCP/RCP after reflection.